\documentstyle[12pt,epsfig]{article}

\def\be{\begin{equation}}
\def\ee{\end{equation}}

\def\IP{\hbox{\rm I\kern -1.6pt{\rm P}}}
\def\IC{{\hbox{\rm C\kern-.58em{\raise.53ex\hbox{$\scriptscriptstyle|$}}
    \kern-.55em{\raise.53ex\hbox{$\scriptscriptstyle|$}} }}}
\def\IN{\hbox{I\kern-.2em\hbox{N}}}
\def\IR{\hbox{\rm I\kern-.2em\hbox{\rm R}}}
\def\ZZ{\hbox{{\rm Z}\kern-.3em{\rm Z}}}
\def\IT{\hbox{\rm T\kern-.38em{\raise.415ex\hbox{$\scriptstyle|$}} }}

\begin{document}

\title{Dynamics of a Massive Piston in an Ideal Gas:  Oscillatory Motion
and Approach to Equilibrium}
\author{N. Chernov$^{1,3}$, J.~L.~Lebowitz$^{2,3}$}
\date{\today}
\maketitle

\begin{center}
{\it Dedicated to Robert Dorfman on the occasion of his 65$^{\,\rm
th}$ birthday}
\end{center}

\begin{abstract}
We study numerically and theoretically (on a heuristic level) the
time evolution of a gas confined to a cube of size $L^3$ divided
into two parts by a piston with mass $M_L \sim L^2$ which can only
move in the $x$-direction. Starting with a uniform
``double-peaked'' (non Maxwellian) distribution of the gas and a
stationary piston, we find that (a) after an initial quiescent
period the system becomes unstable and the piston performs a
damped oscillatory motion, and (b) there is a thermalization of
the system leading to a Maxwellian distribution of the gas
velocities. The time of the onset of the instability appears to
grow like $L \log L$ while the relaxation time to the Maxwellian
grows like $L^{7/2}$. \footnotetext[1]{Department of Mathematics,
University of Alabama at Birmingham, Alabama 35294}
\footnotetext[2]{Department of Mathematics, Rutgers University,
New Jersey 08854} \footnotetext[3]{Current address: Institute for
Advanced Study, Princeton, NJ 08540}
\end{abstract}

\renewcommand{\theequation}{\arabic{section}.\arabic{equation}}

\section{Introduction}
\label{secI} \setcounter{equation}{0}

The time evolution of a gas filled container divided by a massive
piston is an old problem which has attracted much attention
recently from both a conceptual and computational point of view,
\cite{Li,G,KBM}. While we do not believe that there is any
``paradox'' associated with this problem, the basic mechanism of
energy transport across the piston from the hot side to the cold
side which are at equal pressure but different temperatures, was
already described for an idealized version, by one of us in 1959
\cite{L1}, many intriguing questions remain.  Not surprisingly the
microscopic motion of the piston, and thus the time evolution of
the system from an initial, far from equilibrium state, to the
final true equilibrium state is not easy to compute analytically
or even numerically for large systems \cite{KBM}.
It is therefore interesting to consider the
greatly simplified case when the gas particles only interact via
collisions with the massive piston \cite{L1,GF,GP}. This serves,
among other things, as a model for the approach to equilibrium in
such a ``weakly'' interacting macroscopic system.

Here we carry out studies on this system under the ``simplest''
initial conditions. The positions and velocities of gas particles
of unit mass, in an insulated box of size $L^3$, are picked, at
$t=0$, from a Poisson process with some density function, i.e.\
they are assumed to be independent random variables with a given
double peaked velocity distribution $p(v) = p(-v)$
(see (\ref{0cutoff})) and a constant spatial
density throughout the box. The piston, which acts as a single
particle with area $L^2$ and mass $M_L \sim L^2$, is then
released at the position $X(0)=L/2$ with zero velocity $V(0)=0$.

It might be thought that the random fluctuations of the piston's
velocity due to the initial randomness in the positions and
velocities of the gas particles will be relatively small when $L$
is large (the number of gas particles growing like $L^3$) and
vanish in an appropriate hydrodynamic scaling limit (as
$L\to\infty$) of time, space and piston mass.  In fact, some
rigorous results in this direction were obtained in \cite{LPS} and
\cite{CLS} for a class of initial conditions more general than
those considered here. It is proven in those papers that for those
initial distributions the dynamics of the piston, in the
hydrodynamic scaling limit, is governed by deterministic equations
for as long as no particle collides with the piston more than
twice. After that period, however, we could not control the random
fluctuations in the particle configuration anymore. It is clear,
though, that for an initial state which is spatially uniform and
has the same velocity distribution on both sides of the piston,
the deterministic macroscopic evolution would predict that the
position of the piston remains stationary forever. On the other
hand, when the initial velocity distribution of the particles is
not Maxwellian, the system is not at thermal equilibrium, and we
expect it to somehow evolve toward equilibrium for almost any
initial configuration (with respect to the Liouville measure on
the energy surface). This is indeed what we found for every
initial distribution. The path to the Maxwellin taken by the
system turned out however to be quite sensitive to the initial
distribution.

Here we present the results of detailed numerical simulations for
one particular initial density $p(v)$ that vanishes unless
$0.5\leq |v|\leq 1$. We found that the quiescent state becomes
unstable after a few (5--10) recollisions of each gas particle
with the piston. The system then develops large (on a macroscopic
scale) oscillatory motions which last for a long time.  They look
like smooth harmonic oscillations with an (almost) constant
amplitude and piston speed comparable to that of the gas
particles. Eventually, though, the oscillations dampen and vanish,
and the system approaches a stable equilibrium state with constant
gas density and Maxwellian velocity distribution\footnote{The
latter is of course expected on general grounds for, as pointed
out by Boltzmann: ``[the Maxwell distribution] is characterized by
the fact that by far the largest number of possible velocity
distributions have the characteristic properties of the Maxwell
distribution, and compared to these there are only a relatively
small number of possible distributions that deviate significantly
form Maxwell's.  Whereas Zermelo says that the number of states
that finally lead to the Maxwellian state is small compared to all
possible states, I assert on the contrary that by far the largest
number of possible states are ``Maxwellian'' and that the number
that deviate from the Maxwellian state is vanishingly small'',
\cite{B}, see also \cite{L2}.}.

Our main conclusions from the numerical investigations, which we
also ``derive'' heuristically, is that 1) while the dynamics are
non chaotic in the technical sense of there being no positive
Lyapunov exponents, there are for typical initial
conditions\footnote{Typical here means of probability converging
to one (presumably, exponentially fast) as $N\to\infty$ with
respect to the invariant Liouville measure. There are clearly some
exceptional initial configurations (e.g., those with a complete
symmetry about $x=L/2$ with opposite velocities) when $V(t)=0$ at
all times and so the speed of gas particles never changes. We
expect (for the reasons given by Boltzman \cite{B}) all such
initial states to be unstable with respect to generic small
perturbations.} enough interactions to bring the system to
equilibrium along ``interesting'' nontrivial pathways, and 2) that
the time of onset of the instability grows with $L$ as $L \log L$.
This means that even on the hydrodynamical scale $\tau = t/L$, the
deterministic behavior in which nothing happens on the spatial
scale, $y = x/L$, would remain valid when $L \to \infty$.  In
terms of $\tau$ however this onset time grows only like $\log L$
so that in ``practice'' this behavior extends to macroscopic
systems. We believe that this instability is related to the fact
that the deterministic solutions are unstable for the kind of
$p(v)$ we consider here. For other initial $p(v)$, for which the
deterministic solution is stable, we do not get such large
oscillations but, as expected, we still get an approach to
Maxwellian, see Section~7.

\section{Description of the model}
\label{secDM} \setcounter{equation}{0}

Consider a cubical container $\Lambda_L=[0,L]\times [0,L]\times
[0,L]$ filled with an ideal gas consisting of $N$ particles. The
container is divided into two parts by a wall (piston) orthogonal
to the $x$ axis.  At time $t=0$ the wall is released and then it
can move freely without friction inside $\Lambda_L$ along the
$x$-axis, under the action of elastic collisions with the gas
particles,
each of which has the same fixed mass $m$. Since the piston's area
is $L^2$, we assume its mass $M_L$ to be proportional to $L^2$ and
given by $M_L=bmL^2$ with a fixed constant $b>0$ (we set $m=1$ and
$b=2$ in our numerical simulations).

Since the components of the particle velocities perpendicular to
the $x$-axis play no role in the dynamics of the piston, we may
assume that each particle has only one component of velocity, $v$,
directed along the $x$-axis. Hence, each gas particle can be
specified by a pair $(x_i,v_i)$, where $i=1,\ldots,N$. We shall
take the total number of particles $N$ and the total kinetic
energy of the system proportional to $L^3$; and we are interested
in the behavior of the system for $L \gg 1$.

The initial configuration of gas particles and their velocities
$\{(x_i,v_i)\}_{i=1}^N$ is selected at random with statistics
given by a (two-dimensional) Poisson process on the $x,v$ plane
with a given density, $p(x,v) = L^2\,p(v)$. More precisely, for any
domain $D \subset [0,L] \times \IR^1$ the number of particles in
$D$, i.e. $N_D=\#\{i:\, (x_i,v_i) \in D\}$, has a Poisson
distribution with parameter
$$
      \lambda_D=L^2 \int_D p(v)\, dx\, dv
$$
That is, for each $k\geq 0$,
$$
    P(N_D=k)=\frac{\lambda_D^k}{k!}\, e^{-\lambda_D}
$$
The numbers of particles in nonoverlapping domains are independent. The
function $p(v) = p(-v)$ takes values of order one, and the factor of $L^2$
is simply the cross-sectional area of the container $\Lambda_L$. The piston
is initially at rest at the midpoint, $X(0)=L/2$, $V(0)=0$ and the position
and velocity at time $t$ are denoted by $X$, $0 \leq X \leq L$,
and $V$.

We note that the total number of gas particles $N$ in the
container is random. So are the numbers of particles to the left
and to the right of the piston, call them  $N_-$ and $N_+$,
respectively ($N=N_-+N_+$). The initial total kinetic energy of
the system, ${1 \over 2}\sum mv_i^2$, is also a random variable.
Once chosen, the values of $N_-,N_+$ and the total kinetic energy
$E$ of gas plus piston are (presumably, the only) integrals of
motion.

The gas particles and the piston move freely (with constant
velocity) between elastic collisions of particles with the
piston and the walls. When a particle collides with a wall
at $x=0$ or $x=L$, its velocity simply reverses. If a particle
with velocity $v$ hits the piston whose velocity is $V$, then
their velocities after the collision, call them $v^\prime$ and
$V^\prime$, respectively, are given by
\be
    V^\prime = (1-\varepsilon)V + \varepsilon v
      \label{V'}
\ee
\be
    v^\prime = -(1-\varepsilon)v + (2-\varepsilon)V
      \label{v'}
\ee
where
\be
    \varepsilon = \frac{2m}{M+m} =
    \left ( \frac{2}{bL^2+1}\right)
      \label{varepsMm}
\ee

In order to avoid recollisions of gas particles with the piston,
for at least some initial period of time, we impose a velocity
cutoff\footnote{As explained in Section~7, the choice of $p(v)$
plays an important role in the later time evolution of the
system.}
\be
       p(v)\equiv 0 \ \ \ \ \ \
       {\rm if}\ \ \ \ |v|\geq v_{\max}\ \ \ \
       {\rm or} \ \ \ \ |v|\leq v_{\min}
           \label{0cutoff}
\ee
with some $0<v_{\min}<v_{\max}<\infty$, cf. \cite{CLS,LPS}. Under
these conditions, there will be an initial time interval of length
$O(L)$ during which each particle colliding with the piston will
have to travel to the wall, bounce off it, and then travel back to
the piston before it can hit it again. Therefore, during that
interval, there will be no recollisions of any gas particle with
the piston. We call such an interval of time the zero-recollision
interval. Likewise, if the piston remains slow enough after
recollisions occur, there will be another interval of time of
order $L$ during which each gas particle experiences at most one
recollision with the piston. We call it the one-recollision
interval.

The {\em hydrodynamic limit} is now obtained by rescaling space,
$y=x/L$, $0<y<1$, and time, $\tau=t/L$. In the new space-time
coordinates $y,\tau$ the zero-recollision interval and the
one-recollision interval will be of order one, and we will denote
them by $(0,\tau_0)$ and $(\tau_0,\tau_1)$ respectively.  As
already mentioned, we prove in \cite{CLS},  for a more general
class of initial densities $p(x,v)$ that during the interval
$(0,\tau_1)$ the functions $Y_L(\tau)$ and $W_L(\tau)$
$$
      Y_L(\tau)=X(\tau L)/L,\ \ \ \ \ \ \ \ \ \
      W_L(\tau)=dY_L/d\tau=V(\tau L)
$$
converge uniformly in probability,
as $L\to\infty$, to  some deterministic functions $\bar{Y}(\tau)$ and
$\bar{W}(\tau)$; $\bar{W}$ satisfies a certain algebraic equation and
$d\bar{Y}/d\tau = \bar W$.
In the case considered here,
when the density function $p(x,v)$ is independent
of $x$ and symmetric in $v$, i.e. $p(v)=p(-v)$ for all $x,v$, then
the hydrodynamic evolution is trivial: $\bar{Y}(\tau)\equiv 0.5$
and $\bar{W}(\tau)\equiv 0$ for all $\tau>0$. Hence, $Y_L(\tau)\to
0.5$ and $W_L(\tau)\to 0$ as $L\to\infty$, uniformly for all
$\tau\in(0,\tau_1)$.

In this paper we investigate numerically what
happens to $Y_L(\tau)$ and $W_L(\tau)$ beyond the interval
$(0,\tau_1)$. More precisely, for how long does the stochastic
trajectory $\{Y_L(\tau),W_L(\tau)\}$ remain close to the
deterministic one $\{\bar{W}(\tau),\bar{Y}(\tau)\}$? And what does
it look like in the long run?

\section{Numerical results}
\label{secNR} \setcounter{equation}{0}

In the computer simulations described here and
in Sections 4-6, we set
\be
    p(x,v)=p(v)=\left\{\begin{array}{ll}
    1 & {\rm if}\ \ 0.5\leq |v|\leq 1\\
    0 & {\rm elsewhere}
    \end{array}\right .
        \label{pini}
\ee
so $v_{\min}=0.5$ and $v_{\max}=1$. The $x$ and $v$ coordinates of
all the particles are then independent random variables uniformly
distributed in their ranges $0<x<L$ and $v_{\min}\leq |v|\leq
v_{\max}$. Our computer program first selects $N$ according to the
Poisson law with mean $L^3$, and then generates all $(x_i,v_i)$,
$1\leq i\leq N$, independently according to their uniform
distributions. We used the random number generator described in
\cite{MN}. The parameter $L$ changed in our simulations from
$L=30$ to $L=300$. For $L=300$ the system contains $\approx
L^3=27,000,000$ particles.

Once the initial data is generated randomly, the program computes
the dynamics by using the elastic collision rules (\ref{V'}),
(\ref{v'}). All calculations were performed in double precision,
with coordinates and velocities of all particles stored and
computed individually. We note that memory requirements alone can
be enormous -- the program needs over 430Mb RAM in order to run
the model with $L=300$. Also, the task of determining which
particle is to collide with the piston next is nontrivial. To
avoid a long search through the entire set of $N$ particles after
each collision, we assembled a smaller group of particles located
in a vicinity of the piston, where the search is performed, and
updated this group periodically, as time goes on. The calculations
were done in C++ and HPF (High Performance Fortran) on a Dell
Power Edge machine with Dual 733MHz processors at the University
of Alabama in Birmingham. Our code is available on the web page
referred to in the next section.

Figure 1 presents a typical trajectory of the piston. Here
$L=100$. The position and time are measured in hydrodynamic
variables $Y=X/L$, $0<Y<1$, and $\tau=t/L$.

\begin{figure}[h]
\centering \epsfig{figure=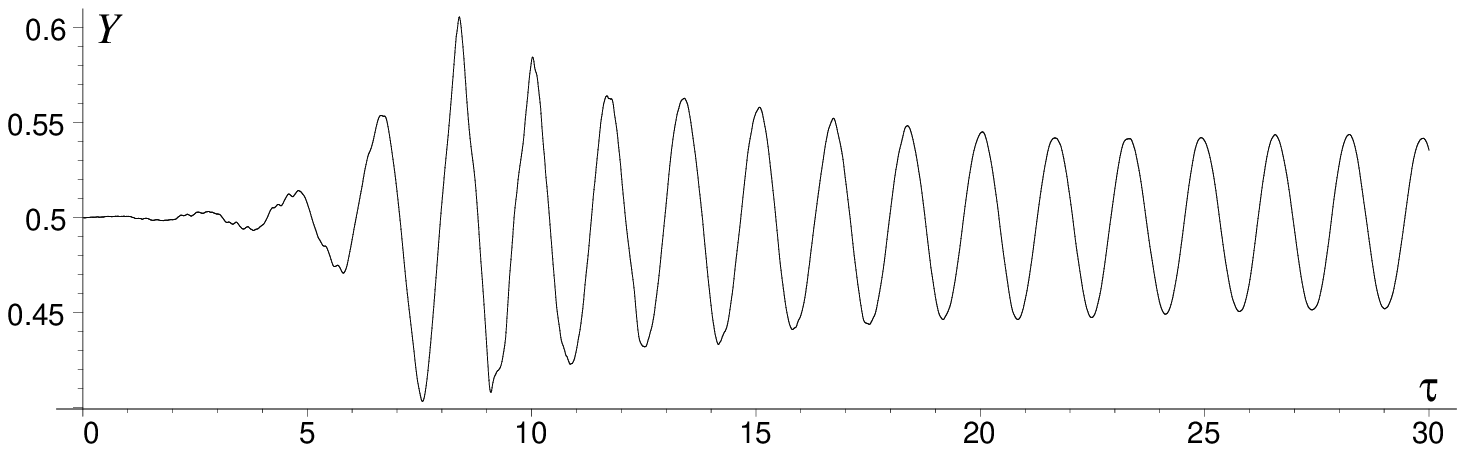}\caption{The piston coordinate
$Y$ as a function of time $\tau$. Here $L=100$, $N_-=500341$,
$N_+=499888$.}
\end{figure}

Initially, the piston barely moves about its equilibrium position:
recall that the hydrodynamic trajectory of the piston is
$\bar{Y}(\tau)\equiv 0.5$ for all $\tau>0$ and that this holds
exactly for $\tau<2$ as $L\to\infty$, \cite{CLS}. Then, at times
$\tau$ between $3$ and $5$, the random vibrations of the piston
grow and become quite visible on the $y$-scale, but for a short
while they look random.  After that the piston starts travelling
back and forth along the $y$ axis, making excursions farther and
farther away from the equilibrium point $y=0.5$. Very soon, at
$\tau = \tau_{\max} \approx 8$, the swinging motion of the piston
reaches its maximum, $(\Delta Y)_{\max}=\max|Y(\tau)-0.5|\approx
0.1$.  Then the oscillations of the piston dampen in size and seem
to stabilize at an amplitude $A\approx 0.04$. At the same time the
trajectory of the piston smoothes out and enters an oscillatory
mode with a period $\tau_{\rm per}\approx 1.63$.

The velocity of the piston $W(\tau)$ follows a similar pattern.
Its random fluctuations grow after $\tau\approx 2.5$, then
it starts swinging up and down, reaches the maximum value of
$W_{\max}=\max|W(\tau)|\approx 0.4$ at time $\tau\approx 9.5$.
After that the oscillations of $W(\tau)$ dampen and seem to
stabilize.

\begin{figure}[h]
\centering \epsfig{figure=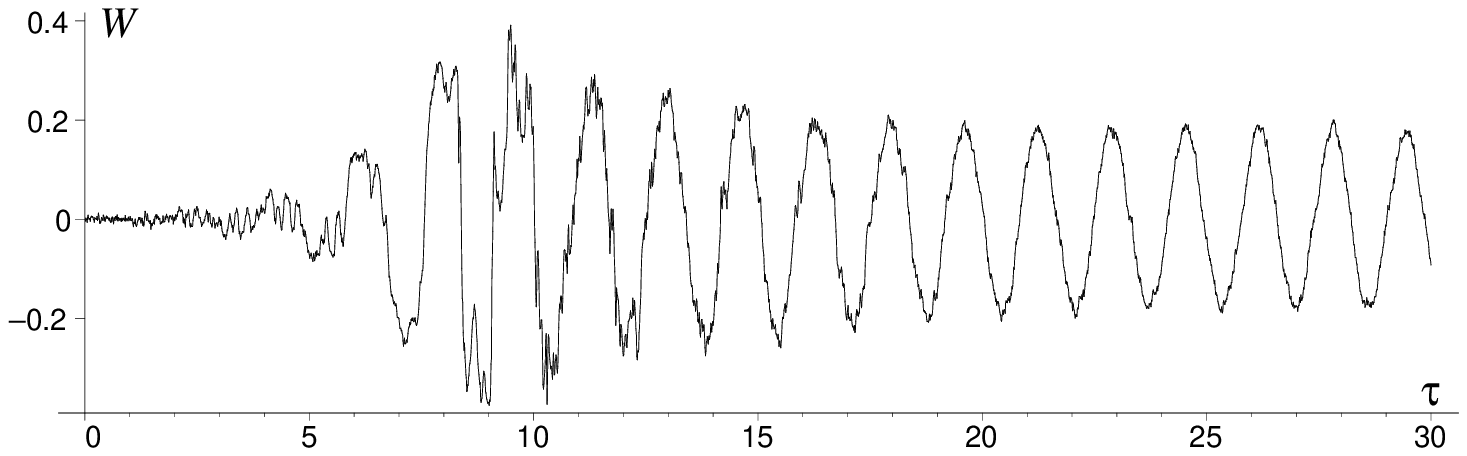}\caption{The piston velocity
$W$ as a function of time $\tau$. The same run as in Fig.~1.}
\end{figure}

Note that the graph of the function $Y(\tau)$ looks much smoother
than that of $W(\tau)$, as would be expected from the fact that
$Y(\tau)$ is the integral of $W(\tau)$.
Interestingly, both functions $Y$ and $W$ smooth out as time goes on.

This cycle of the gas motion in the container between the walls
and the piston continues for a long time with the same period
$\tau_{\rm per} \simeq 1.63$, independent of $L$, but the amplitudes
of both $Y(\tau)$ and $W(\tau)$ are slowly decreasing, see Fig.~3.

\begin{figure}[h]
\centering \epsfig{figure=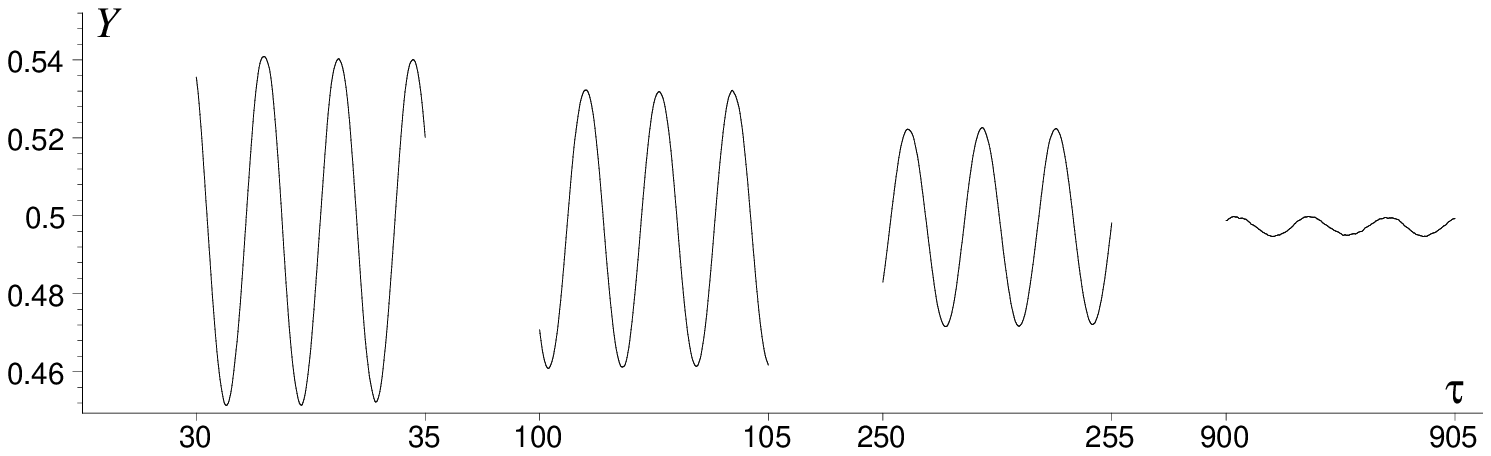}\caption{The piston coordinate
$Y$ during the intervals $(30,35)$, $(100,105)$, $(250,255)$, and
$(900,905)$. The same run as in Fig.~1 and 2.}
\end{figure}

The oscillations of the piston with decaying amplitude can be
described, in the interval $30 <\tau<1000$, approximately by
\be
      Y_1(\tau)\simeq Ae^{-\lambda(\tau-20)}\sin\omega
      (\tau-\alpha)
         \label{Y1}
\ee
with $A = 0.046$ and some constant $\lambda>0$. Correspondingly,
$W_1(\tau)=dY_1/d\tau$ in the same interval $30<\tau<1000$ is
\begin{eqnarray}
      W_1(\tau) &\simeq & -\lambda Y_1+Ae^{-\lambda(\tau-20)}
      \omega\cos\omega (\tau-\alpha)\nonumber\\
      &=& Ae^{-\lambda(\tau-20)}
      [-\lambda\sin\omega (\tau-\alpha)+\omega \cos\omega (\tau-\alpha)]
      \nonumber\\
      &=&
      A_1e^{-\lambda (\tau-20)}\sin\omega (\tau-\beta)
         \label{W1}
\end{eqnarray}
with $A_1=A\sqrt{\omega^2+\lambda^2}$ and some $\beta$ related to
$\alpha$.

To check how well our formula (\ref{Y1}) agrees with the data, we
computed the amplitude $A(\tau)$ as a function of time $\tau$, by
fitting a sine function $Y_0(\tau)=A\sin\omega(\tau-\alpha)$
``locally'', on the interval $(\tau-5,\tau+5)$ for each $\tau$.
Fig.~4 shows $A(\tau)$ on the logarithmic scale, which looks
almost linear on the interval $30<\tau<800$. (After that,
$Y(\tau)$ becomes quite unstable, with already small amplitude
$A(\tau)$ decreasing abruptly, possibly due to the interference
from random fluctuations, so we left that part out.)

\begin{figure}[h]
\centering \epsfig{figure=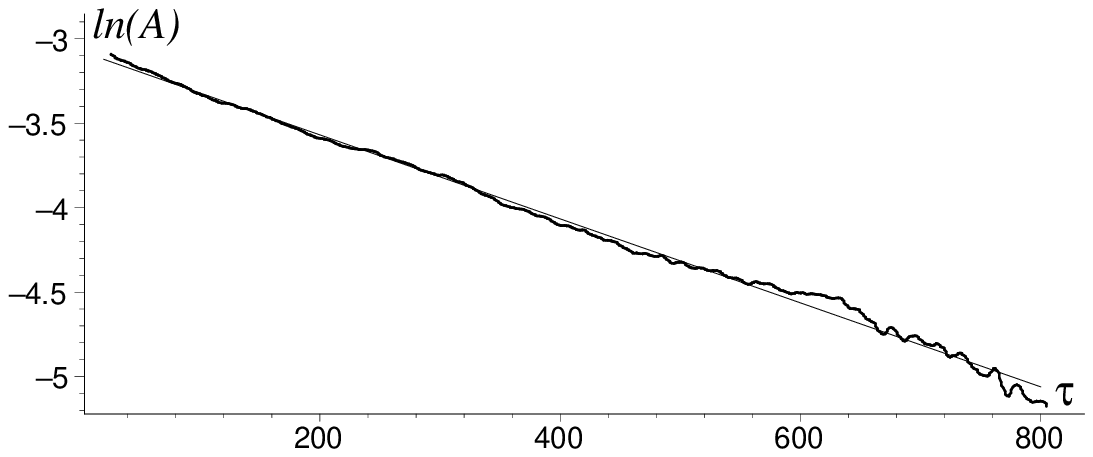}\caption{The amplitude
$A(\tau)$ on the logarithmic scale: experimental curve (bold) and
a linear fit (thin). The same run as the one shown in Fig.~1, 2 and
3.}
\end{figure}

We used the least squares fit to estimate $\lambda=0.00264$ for
the run shown on Figs.~1-4. Since $\lambda$ is small, the
oscillations indeed die out very slowly. The ``half-life'' time
(the time it takes to reduce the amplitude by a factor of two) is
$\tau_{1/2}=\lambda^{-1}\ln 2 \approx 263$. Note that over the
time interval $30<\tau<800$ we only observed the reduction of the
amplitude by a factor of about 10, and the periodic oscillations
were still visible on the plot at $\tau>900$. Also, $\lambda$ and
hence $\tau_{1/2}$ depend on the system size $L$, see below.

\noindent
\section{Dependence of the Time Evolution on System Size}

Many of the characteristics of the piston trajectory described above
($(\Delta Y)_{\max}$, $W_{\max}$, $A$, $\tau_{\rm per}$) appear to
be independent of $L$.
The following table\footnote{The numerical data here
are given primarily for demonstrating
the typical scale of oscillations. They are not
meant to be estimates of physical parameters, so we do not provide
error bars. For $L\leq 200$, where more than one
experimental trajectory was generated, we have estimated
that the accuracy of our numerical values in Table~1
is at least within 10\%.} presents computed
values of all these parameters for different $L$'s. \\

\begin{center}
\begin{tabular}{||r||r|r|c|c|c|r||}
\hline\hline $L$ & $(\Delta Y)_{\max}$ &
$W_{\max}$ & $A$ & $\tau_{\rm per}$ \\ \hline
\hline  30 & 0.114  & 0.40 & 0.042 & 1.70 \\ \hline 50 &
0.121  & 0.39 & 0.045 & 1.65  \\ \hline 80 & 0.105
& 0.37 & 0.042 & 1.65 \\ \hline 100 & 0.100 & 0.34 &
0.041 & 1.64 \\ \hline 120 & 0.122 & 0.39 & 0.041 &
1.62
\\ \hline  150 & 0.111 &  0.37 & 0.045 & 1.62
\\ \hline 200 & 0.102 &  0.37 & 0.037 & 1.62  \\ \hline
250 & 0.100 &  0.41 & 0.042 & 1.64  \\ \hline 300 &
0.122 &  0.42 & 0.045 & 1.65 \\ \hline  \hline
\end{tabular}\vspace*{0.2cm}
\end{center}

\begin{center}
Table 1. Principal characteristics of the piston dynamics.
\end{center}

In each case, we averaged over several experimentally generated
trajectories (for $L=250$ and 300, just one trajectory was used).

There are however other quantities such as $\tau_{\max}$,
$\tau_{1/2}$, and the related $\lambda$, which depend in a systematic way
on $L$. In particular, we estimated numerically that $\tau_{1/2}
\sim L^{1.3}$, hence $\lambda\sim L^{-1.3}$, see Fig.~5. We will
denote $\lambda=\lambda_L$ and discuss it further in Section~6.
We also noticed that in some runs with larger $L$'s (such as $L=150$
and $L=200$) the exponent $\lambda$ changes with time, it is
higher when $\tau<100$ and lower when $\tau>100$.

\begin{figure}[h]
\centering \epsfig{figure=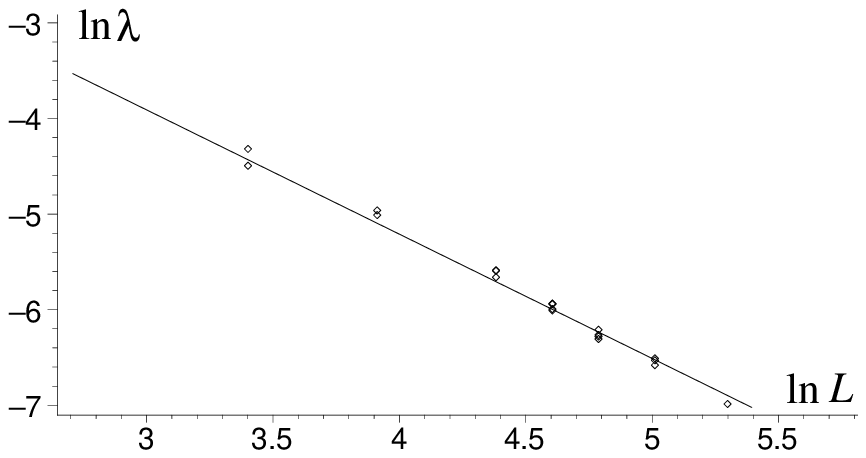}\caption{The value
$\ln\lambda$ as a function of $\ln L$: experimental points
and a linear fit.}
\end{figure}

But most importantly, the time of the largest oscillations
$\tau_{\max}$ and the related time of the onset of the instability
$\tau_c$, see below, seem to slowly grow with $L$, very likely as
$\log L$. To understand this fact, we looked into the mechanism of
the build-up of random fluctuations of the piston position and
velocity, which eventually result in their large nearly harmonic
oscillations. To this end we plotted the histogram of the
(empirical) density of gas particles in the $y,v$ plane at various
times $0<\tau <30$, see samples in Fig.~6. The initial density (at
time zero) is almost uniform over the domain $0<x<L$ and
$v_{\min}\leq |v|\leq v_{\max}$ (variations in the initial
configuration always exist, because it is generated randomly).
Then, for $0<\tau<1$, the piston experiences random collisions
with particles and acquires a speed of order $M_L^{-1/2}=O(1/L)$,
see \cite{L1,Holley,DGL}. These small fluctuations of the piston
velocity result in changes of the velocities  of the particles
which leave the piston after collision. Thus the outgoing
particles on the right hand side of the piston have velocities in
the interval $(v_{\min}+2W(\tau),v_{\max}+2W(\tau))$ while those
on the left hand side of the piston have velocities in the
interval $(-v_{\min}+2W(\tau),-v_{\max}+2W(\tau))$. Hence, the
region in the $y,v$ plane where the density of the particles is
positive is no longer a rectangle with straight sides, now its
boundaries are curves whose shape nearly repeats the graph of the
randomly evolving piston velocity $W(\tau)$. While the variations
of $O(1/L)$ of these boundary curves may seem small, it is crucial
that on opposite sides of the piston they go in opposite
directions. Indeed, when $W(\tau)>0$, then the outgoing particles
on the right hand side accelerate and those on the left hand side
slow down. When $W(\tau)<0$ the opposite happens.

Next, the particles that have collided with the piston travel to
the wall and come back to the piston. Now their densities are less
regular than they were initially -- the regions in the $x,v$ plane
where the density is positive, are curvilinear domains. When they
hit the piston, they shake it back and forth more forcefully than
before, because the velocities of the incoming particles on the
opposite sides of the piston are now negatively correlated. When
particles on the right hand side are fast, those on the left hand
side are slow, and vice versa. This produces a resonance-type
effect destabilizing the piston dramatically and the velocity of
the piston $W(\tau)$ experiences larger fluctuations than before.
The velocities of the newly outgoing particles will again go up
and down in opposite direction, on a greater scale than before.

As time goes on, the above phenomenon repeats over and over, with
larger and larger fluctuations of the gas and piston velocities,
until the distribution of gas particles completely breaks down. At
times $\tau\sim 10$, two large clusters of particles are formed,
one on each side of the piston. When one cluster bombards the
piston, the other moves away from it and hits the wall, then they
exchange their roles. The clusters have sizes of about 0.3--0.5 in the
$y$ direction and the particle velocities range from about 0.2 to
just over 1. The average velocity is about 0.5--0.6 and so the clusters
hammer the piston periodically with period 1.6--2.0,
which is close to the experimentally determined period of piston
oscillations, see above.

\begin{figure}[h]
\centering \epsfig{figure=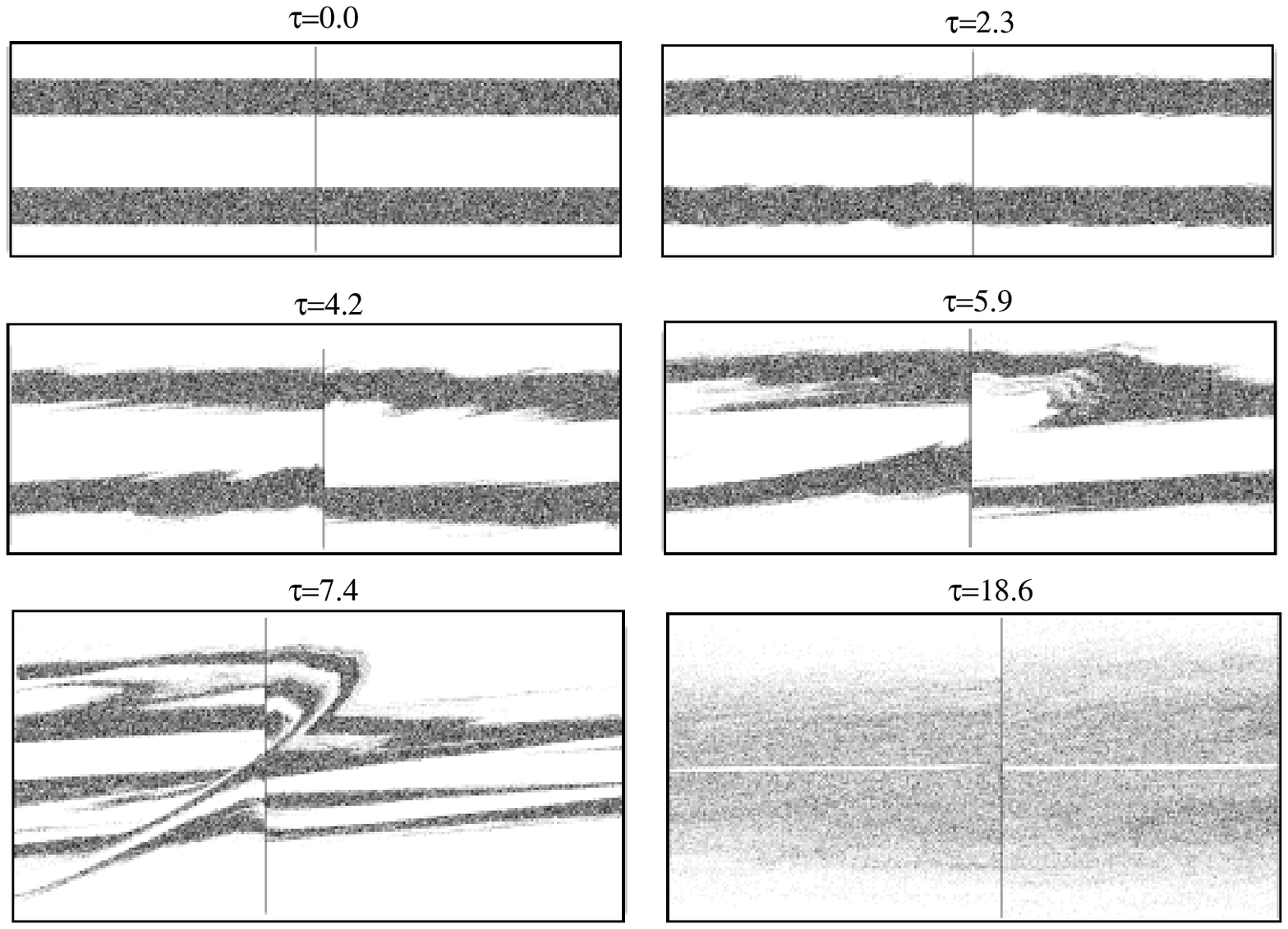}\caption{Six snapshots of the
empirical gas density (in the $x,v$ plane)
at times $\tau=0$, 2.3, 4.2, 5.9, 7.4 and 18.6.}
\end{figure}

Fig.~6 shows six snapshots of the empirical density of gas
particles taken at different times. At $\tau=0$ the gas fills
(almost uniformly) two rectangles $\{(y,v):\ 0.5<|v|<1,\ 0<y<1\}$.
At $\tau=2.3$ one can see some ripples on the boundaries of these
rectangles. At time $\tau=4.2$ the irregularities grow and at
$\tau=5.9$ the rectangular shape is broken down. Two large
clusters of particles are formed, both appear in the upper
half-plane $v>0$, i.e.\ at that time both clusters move to the
right (one toward the piston, the other away from it). Later the
density undergoes strange formations ($\tau=7.4$) but eventually
smoothes out and enters a slow process of convergence to
Maxwellian ($\tau=18.6$) described below. Note a narrow white line
around $v=0$, meaning the total lack of very slow particles at
time $\tau=18.6$.

A longer sequence of snapshots at times $0<\tau<30$ is posted
on the web page
$$
   {\rm www.math.uab.edu/chernov/piston/pictures/piston.html}
$$
It gives a spectacular view of the entire system
evolution.

The above analysis may suggest that the fluctuations of the piston
velocity roughly increase by a constant factor during each time
interval of length one. Indeed, initial random fluctuations
$W_a\sim O(1/L)$ result in additional changes of velocities of
outgoing particles by $2W_a$. When those particles come back to
the piston (in time $\Delta \tau\approx 1$), they kick its
velocity to the level of $2W_a$. Then the newly outgoing particles
acquire an additional velocity $4W_a$, etc. Over each time
interval of length one the fluctuations double in size. This is an
obvious oversimplification of the real dynamics, but it leads to a
reasonable conjecture
\be
    W_a(\tau) \approx \frac{C\, R^{\tau}}{L}
      \label{Wa}
\ee
where $W_a(\tau)$ are typical fluctuations of the piston velocity
at time $\tau$ and $C,R>0$ are constants.

\begin{figure}[h]
\centering \epsfig{figure=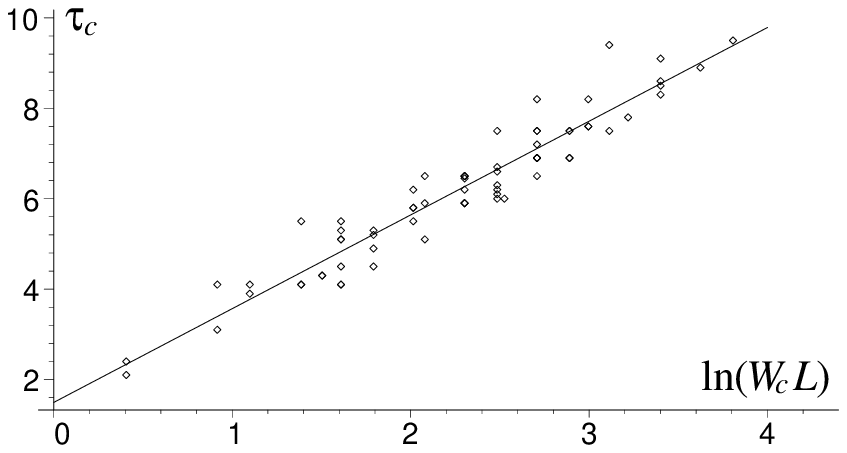}\caption{The value $\tau_c$ as
a function of $\ln (W_cL)$: experimental points and a linear fit.}
\end{figure}

We tested the above formula numerically as follows. Let $W_c>0$
be some preset critical value of the piston speed and
$\tau_c=\inf\{\tau>0:\, |W(\tau)|\geq W_c\}$ the (random) time
when $W_c$ is first reached. This time plays the role of the
``onset'' of large fluctuations of the piston velocity. One would
expect, based on (\ref{Wa}) that
\be
      \tau_c \approx \ln (W_cL/C)/\ln R
        \label{tauc}
\ee
i.e. $\tau_c$ grows as $\ln L$ when $L$ increases.

We found $\tau_c$ experimentally for $W_c=0.1$ and $W_c=0.15$ and
checked that (\ref{tauc}) agreed well with the data, see Fig.~7.
By the least squares fit we estimated $C=0.45$ and $R=1.6$.

\section{Approach to equilibrium}
\label{secRE} \setcounter{equation}{0}

We examined the convergence of the velocity
distribution of gas particles to a Maxwellian. At any given time
$\tau>0$, let
$$
   F_{\tau}(u)=\#\{i:\, v_i<u\}/N
$$
be the empirical (cumulative) distribution function of particle
velocities. At equilibrium, it should be close to the normal
distribution function $\Phi(x)$. As a measure of their closeness, we used
the supremum of the difference
$$
           D_{\tau}=\sup_{-\infty<u<\infty}|F_{\tau}(u)-\Phi(u)|
$$
Initially, $D_0\approx 0.245$ for our choice of $p(v)$. One can
expect that $D_{\tau}$ converges to 0 as $\tau$ grows when $N$ is
large. In fact, if the velocities $v_i$ were independent and had
Maxwellian distribution (which it would be in true statistical
equilibrium), then $D_{\tau}$ would be of order $O(1/\sqrt{N})$,
and the product $D_{\tau}\sqrt{N}$ would have a certain limit
distribution, see the theory of the Kolmogorov-Smirnov
statistical test \cite{Lu}. In particular, it is known that for
a Maxwellian the probability $P(D_{\tau}\sqrt{N}>1)\approx 0.2$. Based on this, we
define the time of convergence to equilibrium by
\be
           \tau_{\rm eq}=\inf\{\tau>0:\, D_{\tau}\sqrt{N}<1\}
              \label{taueq}
\ee
Here the constant 1 as a critical value is chosen arbitrarily. We
estimated $\tau_{\rm eq}$ for various $L$'s and found that
$\tau_{\rm eq}\approx aL^b$ with some constants $a,b>0$. By a
least squares fit we found $a=0.18$ and $b=2.47$, see Fig.~8.
(Note that the accuracy of our experimental data seems to increase
with $L$, as the points are getting closer to each other and to
the fitting line for larger $L$'s on Fig.~8.)

\begin{figure}[h]
\centering \epsfig{figure=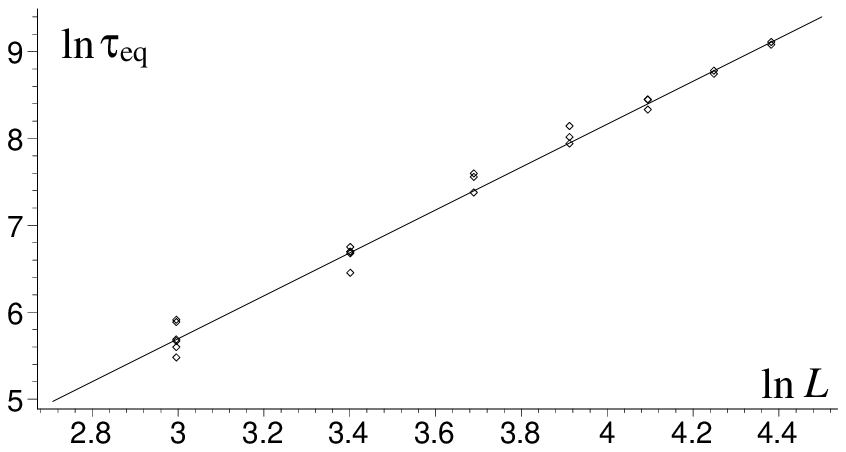}\caption{The value
$\ln\tau_{\rm eq}$ as a function of $\ln L$: experimental points
and a linear fit.}
\end{figure}

The plot of the product $S=D_{\tau}\sqrt{N}$ versus $\tau$ is
given on Fig.~9 (for a particular run with $L=40$). It shows that,
after an initial sharp drop for $0<\tau<20$, the statistic $S$
decreases exponentially in $\tau$. Another commonly used statistic
to measure closeness to Maxwellian is
$$
           S'=3-\frac{M_4}{M_2^2}\
$$
where $M_2$ and $M_4$ are the second and the fourth sample moments
of the empirical velocity distribution, respectively. Fig.~9 shows
that $S'$ converges to zero in a similar manner (for the same run
with $L=40$).\\

\begin{figure}[h]
\centering \epsfig{figure=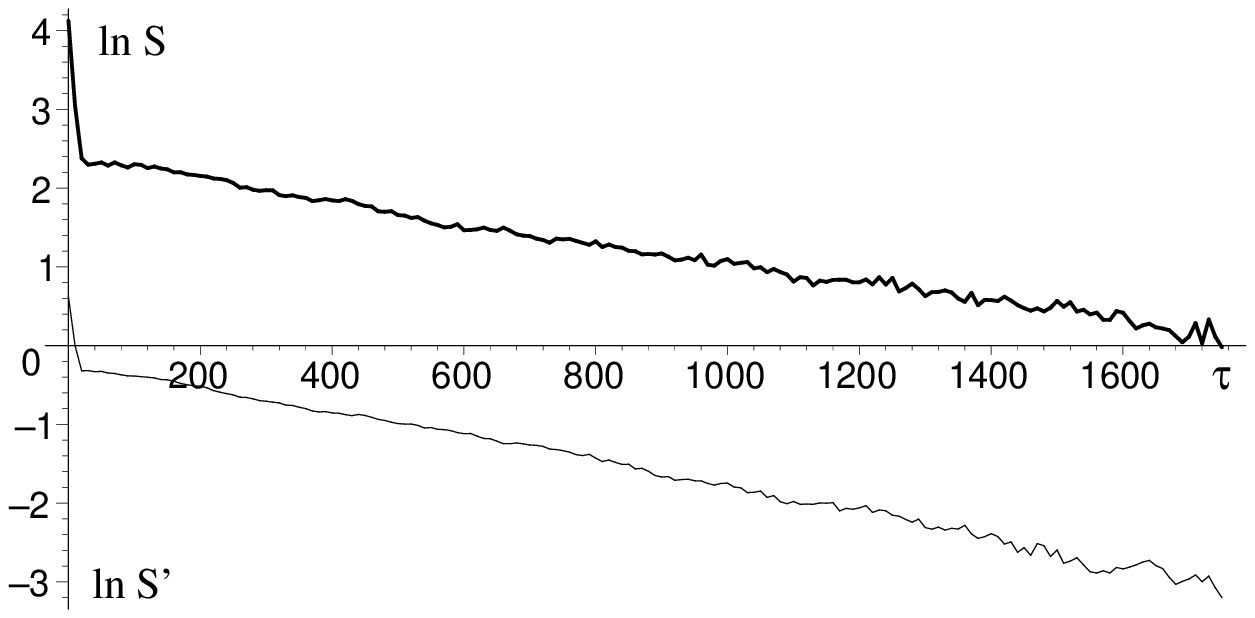}\caption{$\ln S$ (thick line)
and $\ln S'$ (thin line) as functions of $\tau$.}
\end{figure}

\noindent
{\em Theoretical Considerations}

As we noted in Section~\ref{secDM}, the total energy $E$ and the
numbers of particles in the left and right compartments
($N_-$ and $N_+$) are integrals of motion. With these quantities
fixed, the  model can be reduced to a billiard system in a
high-dimensional polyhedron by standard techniques, as we show
next.

Let $\{x_i\}$, $i=1,\ldots,N_+$, denote the $x$-coordinates of the
particles to the right of the piston, and  $\{x_i\}$,
$i=-1,\ldots,-N_-$, those to the left of it (ordered arbitrarily).
Put $x_0=X\sqrt{M}$, where $X$ is the coordinate of the piston and
$M$ is its mass. Then the configuration space of the system (in
the coordinates $x_i$, $-N_-\leq i\leq N_+$) is a polyhedron
$Q\subset \IR^{N+1}$ (recall that $N=N_-+N_+$) defined by
inequalities
$$
   0\leq x_{-N_-},\ldots,x_{-1}\leq x_0/\sqrt{M} \leq x_1,\ldots,x_{N_+}\leq L
$$
It is known that the dynamics of our mechanical system
(``gas+piston'') corresponds to the billiard dynamics in $Q$, see
\cite{CFS}. That is, the configuration point ${\bf q}\in Q$ moves
freely and experiences specular reflections at the boundary
$\partial Q$. The velocity vector
$$
    {\bf p}=\dot{\bf q}=\{v_{-N_-},\ldots,v_{-1},V\sqrt{M},v_1,\ldots, v_{N_+}\}
$$
has constant length, since $\|{\bf p}\|^2=2E=\,$const. Therefore,
the phase space of the billiard system is ${\cal M}=Q\times
S^N_{\rho}$ where $S^N_{\rho}$ is the $N$-dimensional sphere of
radius $\rho=\sqrt{2E}$.

The billiard system has a natural equilibrium state given by the
Liouville measure $\mu$ on $\cal M$, which is the product of a
uniform measure on the polyhedron $Q$ and a uniform (Lebesgue)
measure on the sphere $S^N_{\rho}$, i.e. $d\mu=dq\, dp$. The
properties of billiard dynamics depend heavily on the curvature of
the boundary $\partial Q$. In our case $Q$ is a polyhedron, hence
its boundary consists of flat sides with zero curvature. A
prototype of such systems is billiard in a polygon. It is well
known that (see, e.g., \cite{C})

\medskip
\noindent{\bf Fact}. For billiards in polygons and polyhedra (and
hence, for our mechanical model of a piston in the ideal gas) all
Lyapunov exponents vanish, and so does the Kolmogorov-Sinai
entropy.
\medskip

Systems with zero Lyapunov exponents and zero entropy are not
regarded as chaotic, but they still may be ergodic. In fact,
billiards in generic polygons {\em are} ergodic \cite{KMS}.
Moreover, for many nonergodic polygons, the phase space is
foliated by invariant subsurfaces on which the dynamics is
ergodic.

Even though there are no similar results, to our knowledge, for
billiards in high-dimensional polyhedra, one can expect that they,
too, have similar properties. That is, they are generically
ergodic or become ergodic after trivial reductions. In our case,
the billiard in $Q$ is, perhaps, ergodic for typical values of
$M$, or else the phase space is foliated by invariant submanifolds
on which the dynamics is ergodic, and that those submanifolds fill
$\cal M$ pretty densely. In the latter case, one would hardly
distinguish experimentally between such a nonergodic system and a
truly ergodic one.

Hence, we can assume that our system is ergodic or very close to
ergodic in the above sense. Then almost every trajectory
eventually behaves according to the invariant measure $\mu$,
independently of the initial state. In particular, for any initial
gas density and velocity distribution (given by the function
$p(x,v)$, see Section~\ref{secDM}) the hydrodynamic regime for a
finite $L$ is only valid on a finite interval of time --
eventually the system will relax to equilibrium.  We expect in
fact that in terms of the ``macroscopic'' variables, say, the one
particle distribution function, the system will relax to an
effective equilibrium, as defined by (\ref{taueq}) in terms of
$\tau_{\rm eq}$, which is much smaller than the exponentially long
time (in $L$) required for the ergodic theorem. So the real
question is how does this time depend on $L$. According to our
earlier discussion $\tau_c \sim \log L$ and $\tau_{\rm eq} \sim
L^{5/2}$.

At equilibrium, the distribution of coordinates $x_i$ and
velocities $v_i$ are determined by the Liouville measure $\mu$,
which is uniform in the phase space. Physically interesting (and
only observable) are its marginal measures, i.e. projections, on
lower-dimensional subspaces. The marginal measures of the velocities are
normal (Gaussian) for large $N$ (as $N\to\infty$).

In particular, each individual velocity $v_i$ converges in law to
a Maxwellian (i.e., normal) distribution with zero mean and
variance $2E/N=\,$const. The same holds for the ``piston''
component of the velocity, $\dot{x}_0=V\sqrt{M}$, hence the piston
velocity $V$ will be normally distributed with zero mean and
standard deviation const$/\sqrt{M}=\,$const$/L$, as $L\to\infty$.
In our case $V$ has standard deviation $\sqrt{7/24}/L\approx
0.5/L$. This conclusion agrees well with our numerical
data\footnote{It also allows us to estimate, in a peculiar way,
the time of convergence to equilibrium, $\tau_{\rm eq}$, discussed
earlier. Assume all the initial velocities are of order one, and
note that the largest Maxwellian velocities are of order
$\sqrt{N}=L^{3/2}$. Since each collision with the piston adds
$O(1/L)$ to a particle's velocity, it takes $\sim L^{5/2}$
collisions to reach the maximum. An amazing agreement with the
estimate $\tau_{\rm eq}\sim L^{2.47}$ reported near Fig.~8.}.

The equilibrium distribution of the piston coordinate $X$ is also
determined by the projection of the uniform measure $dq$ on $Q$
onto the $x_0$ axis. Before we do that, let us get rid of $M$ in
the definition of both $Q$ and $x_0$. A simple change of variable
$X=x_0/\sqrt{M}$ allows us to redefine $Q$ by
$$
   0\leq x_{-N_-},\ldots,x_{-1}\leq X \leq x_1,\ldots,x_{N_+}\leq L
$$
Furthermore, rescaling $Y=X/L$ and $y_i=x_i/L$ gives a new,
simpler, definition of $Q$:
$$
   0\leq y_{-N_-},\ldots,y_{-1}\leq Y \leq y_1,\ldots,y_{N_+}\leq 1
$$
``Integrating away'' the variables $y_i$ yields the
following equilibrium density for $Y$:
$$
            f(Y)=c\, Y^{N_-}(1-Y)^{N_+}
$$
for $0<Y<1$, where $c$ is  the normalizing factor that can be
computed explicitly. Put $z=(Y-0.5)\sqrt{8K}$, then the density of
$z$ is given asymptotically by $f(z) \approx
\frac{1}{\sqrt{2\pi}}\, e^{-z^2/2}$. Hence, $Y$ is asymptotically
gaussian with mean $0.5$ and variance $(4N)^{-1}=(4L^3)^{-1}$.
Therefore, in equilibrium
$$
         |Y-0.5|\sim \frac{1}{2L\sqrt{L}}\sim\frac{1}{2\sqrt{N}}
$$
and the probability of observing larger fluctuations is
exponentially small in $N$.
Note that fact is independent of the piston mass, as it has to be for
an equilibrium (classical) system.  This is consistent
with our observation of the piston coordinate reaching
a positive constant value, $|Y-0.5|\approx 0.1$,
during a short time interval $0<\tau<20$, since
our initial conditions were selected according to the
density function $p(x,v)$ which has probability
exponentially small in $N$.

\section{Remarks}
\label{secR} \setcounter{equation}{0}

1. We checked the accuracy of our computer program in various
ways. A simple one was based on the effect of round-offs on the
energy of the system: the total energy was found to be practically
constant over the whole period $0<\tau<1000$. A more sensitive
test of the accuracy of the computations consists in using the
time-reversal symmetry of the dynamics. Suppose at some time $\bar
\tau>0$ the velocities of all gas particles and the piston are
reversed ($v_i\to -v_i$). Then the system is supposed to trace
back its past trajectory and arrive to its initial state with all
velocities reversed at time $2\bar \tau$. We verified this
property numerically for various $\bar \tau<50$ and always found
that the system did repeat its past trajectory in the sense that
the graphs of $Y(\tau)$ and $W(\tau)$ over the interval $(\bar
\tau,2\bar \tau)$ looked like perfect mirror images of the
corresponding graphs over the interval $(0,\bar \tau)$. Therefore,
one can assume that round off errors remain negligibly small
during times $\tau<50$. However, this accuracy test failed for
larger times, $\bar \tau>100$, indicating that the system then
``loses memory'' of its initial state due to round-offs. This does
not seem to affect the overall picture, though. As yet another
test, we carried out some computations with single rather than
double precision, and found that the changes were little. Still,
some estimates requiring long runs, such as the equilibrium time
$\tau_{\rm eq}$ on Fig.~8, might not be very accurate.\\

2. We tested the dependence of the piston oscillations
on the $b$ factor involved in the formula
$M_L=bmL^2$. Remember that we set $b=2$
in our main experiments. When we changed it to $b=20$
(this made the piston 10 times heavier), then the oscillations
started slightly later and their amplitude was slightly larger,
but otherwise the picture was
very much the same. When we changed $b$ to 0.2 (and
this made the piston 10 times lighter), then the oscillations
started at about the same time as for $b=2$, but they
dampened somewhat faster.

We also tried to change the piston mass $M_L$ even more
drastically. When we set it to $L^3$ (insteas of $L^2$), it
appeared that oscillations started only after a very long initial
quiescent period. But we did not examine this fact in detail, see
\cite{GPL}. On the contrary, when we set the piston mass $M_L$ to
$L$ (instead of $L^2$), large oscillations started very soon, but
very quickly dampened and disappeared.\\

3. We note that Eq.\ (\ref{W1}) describes the time evolution
of a damped harmonic oscillator. Accepting (\ref{W1})
over some time range, say, $\tau\in [30,800]$ for $L=100$, we can
then look at the ``inverse problem'' of finding the effective
spring constant and damping coefficient.

Using the original variables, $t$ and $X_L(t)$,
we write
$$
     M_L\frac{d^2X_L}{dt^2}+K_L(X_L-L/2)
     +\eta_L\frac{dX_L}{dt}=0
$$
which has the solution $X_L-L/2\sim e^{\alpha t}$ with
$$
   \alpha=-\frac{\eta_L}{2M_L}\pm
   i\,\sqrt{\frac{K_L}{M_L}}\,
   \left [1-\frac{\eta_L^2}{4K_LM_L}\right ]^{1/2}
$$
This yields, to the lowest order in $\eta_L^2/4K_LM_L$, remembering
that $M_L=2L^2$ and that $\omega= 2\pi/\tau_{\rm per}$, with $\tau_{\rm
per}=t_{\rm per}/L \approx 1.63$, $K_L\approx 8\pi^2/\tau_{\rm per}^2$,
i.e. the effective ``restoring force'' $K_L$ is independent of $L$. On
the other hand, the damping coefficient is $\eta_L=4L\lambda_L$, were
$\lambda_L$ was found experimentally to decrease as $\lambda_L\sim
L^{-\gamma}$ with $\gamma=1.3$, see Section~4. Hence $\eta_L\sim
L^{1-\gamma}=O(L^{-0.3})$, i.e. the damping gets weaker as
$L\to\infty$.

\section{Discussion}
\label{secD} \setcounter{equation}{0}

We have presented here numerical results concerning the time
evolution of a system with many degrees of freedom (up to
$27\times 10^6$) one of which, the position of the piston, plays a
very special role. Starting with the particular initial particle
distribution given by (\ref{pini}) we found two striking features
of the evolution: (1) the velocity distribution of the particles
approaches a Maxwellian, i.e. the system goes toward thermal
equilibrium and (2) the time evolution of the piston followed
closely, after some initial period, that of a damped harmonic
oscillator over an extended time interval, with initial
oscillations as large as 1/10 of the system size.

As already noted, we expect from general considerations \cite{B} that
(1) should be true for any initial density $p(x,v)$ provided the number
of particles $N$ ($\sim L^3$) is large enough and one waits long
enough. But what about (2)? Clearly, if we choose for $p(|v|)$ a
Maxwellian, we expect only thermal fluctuations of order $O(L^{-3/2})$
in the position of the piston. This is indeed what we found
numerically. For other, non-maxwellian, initial densities the question
turns out to be far from trivial. We are currently working with more
general initial densities and will report results in a separate paper
\cite{CCLP}. Below we outline our program and mention some preliminary
findings.

For simplicity, we assume that $p(x,v)=p(|v|)$ and $X(0)=L/2$,
$V(0)=0$. In this case there is no apriori bias for the piston to
move at all, and as already noted, the hydrodynamical
(deterministic) equations, see \cite{CLS}, predict that, in the
limit $L\to\infty$, the system would remain frozen in the initial
state: $Y(\tau)\equiv Y(0)=0.5$ and $W(\tau)\equiv W(0)=0$ for all
$\tau>0$ (in the variables $y$ and $\tau$, see Section~2). The
density $p(y,v,t)$ will also remain constant in time. But what
about the particle system with a large but finite $L$? How will
the piston and the gas behave while the particle velocities make
their way to a Maxwellian?

To answer this question we note that since the initial
configuration of particles is generated randomly from a Poisson
process with the density $p(y,v,0)=p(|v|)$, the actual (empirical)
density of the particles, such as the one shown on Fig.~6 at
$\tau=0$, does not exactly coincide with $p(y,v,0)$. Random
fluctuations of the empirical density are typically of order
$O(1/L)$. Hence, the actual initial distribution of particles can
be thought of as a small perturbation of the function $p(|v|)$ and
can be written as $p(|v|)+\varepsilon p_1(y,v)$ with
$\varepsilon=1/L$ and some (random) function $p_1(y,v)$ of order
one.

Now, we conjecture that for large $L$ the evolution of the
particle system closely follows the solutions of the
hydrodynamical equations derived in \cite{CLS,LPS} with a
perturbed initial density $p(|v|)+\varepsilon p_1(y,v)$, rather
than the stationary solution corresponding to the unperturbed
density $p(|v|)$. In particular, if the latter solution is
unstable, then small perturbations grow exponentially in time (in
the units of $\tau=t/L$, of course), hence the system can be
destabilized in $\tau\sim -\log \varepsilon=\log L$ units of time.
This would be in agreement with our analysis and numerical
estimates in the end of Section~4. Hence, the instability of the
hydrodynamical equations becomes a key issue.

When we were finishing the present paper, we received a message
from E.~Caglioti and E.~Presutti who (a) proved that the
hydrodynamical equations are stable when $p(|v|)$ is monotonically
nonincreasing in $|v|$, i.e.\ $p'(|v|)\leq 0$, and (b) suggested
that they might be unstable for our class on non-monotone
$p(|v|)$. We checked the suggestion (b) for our particular density
(\ref{pini}) and found that it was indeed correct; we proved that
small perturbations grow exponentially in $\tau$. Furthermore,
when starting with a perturbed initial density with
$\varepsilon=0.01$, we found numerically that the corresponding
solution of the hydrodynamical equations resembled very well the
evolution of the particle system described here, including large
nearly harmonic oscillations of the piston during the interval
$10<\tau<30$. Further work in this direction is currently under
way \cite{CCLP}.

Conversely, when we simulated a particle dynamics with a
nonincreasing initial density $p(|v|)$ the oscillations
essentially disappeared\footnote{We tried a uniform ``flat''
function given by $p(|v|)=1$ for $|v|\leq v_{\max}$ and a
triangular one $p(|v|)=1-|v|/v_{\max}$.}. On the other hand, the
particle velocity distribution still approached a Maxwellian,
albeit at a somewhat slower pace.

\medskip\noindent
{\bf Acknowledgement}. We thank F.~Bonetto, R.~Dorfman, Ch.~Gruber,
J.~Piasecki, N.~Simanyi, Ya.~Sinai, and V.~Yakhot for many useful
discussions. We thank H.~van~den~Bedem and C.~Lesort for technical
assistance in numerical experiments. N. Chernov was partially supported by
NSF grant DMS-0098788. J. Lebowitz was partially supported by NSF grant
DMR-9813268 and by Air Force grant F49620-01-0154. This work was completed
when the authors stayed at the Institute for Advanced Study with partial
support by NSF grant DMS-9729992.

\end{document}